\documentstyle[11pt,hrd_pasp,twoside,epsf]{article}
\def\edcomment#1{\iffalse\marginpar{\raggedright\sl#1\/}\else\relax\fi}
\reversemarginpar

\begin{document}
\title{GAIA: Understanding the Galaxy}
 \author{Gerard Gilmore}
\affil{Institute of Astronomy, Madingley Rd, Cambridge CB3 0HA UK}

\begin{abstract}

The GAIA Observatory, ESA's Cornerstone~6 mission, addresses the
origin and evolution of our Galaxy, and a host of other scientific
challenges.  GAIA will provide unprecedented positional and radial
velocity measurements with the accuracies needed to produce a
stereoscopic and kinematic census of about one billion stars in our
Galaxy and throughout the Local Group, about one per cent of the
Galactic stellar population.  Combined with astrophysical information
for each star, provided by on-board multi-colour photometry, these
data will have the precision and depth necessary to address the three
key questions which underlie the GAIA science case:\\

When did the stars in the Milky Way form?\\

When and how was the Milky Way assembled?\\

What is the distribution of dark matter in our Galaxy?

\vskip 10pt

The accurate stellar data acquired for this purpose will also have an
enormous impact on all areas of stellar astrophysics, including
luminosity calibrations, structural studies, and the cosmic distance
scale.  Additional scientific products include detection and orbital
classification of tens of thousands of extra-solar planetary systems,
a comprehensive survey of objects ranging from huge numbers of minor
bodies in our Solar System, including near-Earth objects, through
galaxies in the nearby Universe, to some 500\,000 distant
quasars. GAIA will also provide a number of stringent new tests of
general relativity and cosmology.

There are many scientific tasks to optimise GAIA which demand
immediate effort, providing an ideal opportunity to play a major role in
the project, and in the future of astronomy. Those interested in
being part of GAIA activities, or wanting to know more, should contact
Michael Perryman (mperryma@astro.estec.esa.nl) or Gerry Gilmore
(gil@ast.cam.ac.uk) and see {\tt http://astro.estec.esa.nl/GAIA}.

\vskip 10pt You can make your own three-dimensional model of the GAIA
satellite by downloading the instructions and model parts from the WWW
page. To make the model all you need is paper, glue, scissors and a
spent match stick.

\end{abstract}

\section{Introduction}

GAIA is the astrophysics mission selected as Cornerstone~6 in the ESA
science programme. The acronym has many interpretations, with Galactic
Astrophysics through Imaging and Astrometry being most apt.

GAIA builds upon the observational techniques pioneered and proven by
ESA's Hipparcos mission to solve one of the most difficult yet deeply
fundamental challenges in modern astronomy: to create an
extremely precise three-dimensional map of a representative sample of
stars throughout our Galaxy and beyond. In the process, by combining
positional data with complementary radial velocities, GAIA will map
the stellar motions, which encode the origin and subsequent evolution
of the Galaxy.  Through comprehensive photometric classification, GAIA
will provide the detailed physical properties of each star observed:
characterizing their luminosity, temperature, gravity, and elemental
composition.  This massive multi-parameter stellar census will provide
the basic observational data to quantify the origin, structure, and
evolutionary history of our Galaxy, the primary science goal of the
GAIA mission. 

GAIA will achieve this by repeatedly measuring the positions and
multi-colour brightnesses of all objects down to $V=20$~mag. On-board
object detection will ensure that variable stars, supernovae,
transient sources, micro-lensed events, and minor planets will all be
detected and catalogued to this faint limit. Final accuracies of
10~microarcsec at 15~mag, comparable to the diameter of a human hair
at a distance of 1000~km, will provide distances accurate to 10~per
cent as far as the Galactic Centre, 30\,000 light years away. Stellar
motions will be measured even in the Andromeda galaxy.

\section{GAIA: The Scientific Case}

The range of scientific topics which will be addressed by the GAIA
data is vast, covering much of modern astrophysics, and fundamental
physics. In this section we present a few illustrative examples, to
give the flavour of the mission capabilities, with scientific
applications ranging from the Milky Way Galaxy, stellar astrophysics,
Solar System minor bodies, and extra-galactic studies, to fundamental
physics. Further details are available on the GAIA www site {\tt
http://astro.estec.esa.nl/GAIA}.  Documents there contain references
to the original work briefly summarised here, as well as details of
the many other exciting scientific projects which GAIA will address, but
which space precludes discussion of here.

\subsection{Structure and Evolution of the Milky Way Galaxy}
Understanding the Galaxy in which we live is one of the great
intellectual challenges facing modern science. The Milky Way contains
a complex mix of stars, planets, interstellar gas and dust, radiation,
and the ubiquitous dark matter. These components are widely
distributed in age (reflecting their birth rate), in space (reflecting
their birth places and subsequent motions), on orbits (determined by
the gravitational force generated by their own mass), and with
chemical element abundances (determined by the past history of star
formation and gas accretion).  Astrophysics has now developed the
tools to measure these distributions in space, kinematics, and
chemical abundance, and to interpret the distribution functions to
map, and to understand, the formation, structure, evolution, and
future of our Galaxy. This potential understanding is also of profound
significance for quantitative studies of the high-redshift Universe: a
well-studied nearby template underpins analysis of unresolved galaxies
with other facilities, and at other wavelengths.

Understanding the structure and evolution of a galaxy requires three
complementary observational approaches: (i) a census of the contents
of a large, representative, part of the galaxy; (ii) quantification of
the present spatial structure, from distances; (iii) knowledge of the
three-dimensional space motions, to determine the gravitational field
and the stellar orbits. That is, one requires complementary
astrometry, photometry, and radial velocities. Astrometric
measurements uniquely provide model independent distances and
transverse kinematics, and form the base of the cosmic distance scale.
Photometry, with appropriate astrometric and astrophysical
calibration, gives a knowledge of extinction, and hence, combined with
astrometry, provides intrinsic luminosities, spatial distribution
functions, and stellar chemical abundance and age information. Radial
velocities complete the kinematic triad, allowing determination of
gravitational forces, and the distribution of invisible mass. 
The combination of vast continuing ground-based radial velocity projects
and Hipparcos did this for one location in the Milky Way, the
Solar neighbourhood; GAIA will accomplish this for a large fraction of
our Galaxy.

GAIA will measure not only the local kinematics with much improved
accuracy, but the full six-dimensional stellar distribution function
throughout a large part of the Galactic disk. This will allow not only
a determination of the gravitational potential of the Galaxy and its
distribution function, but also reveal how much a given stellar
population deviates from dynamical equilibrium. This in turn will
constrain the formation history of the Galactic disk and its
components, e.g., the past variations of pattern speed and strength of
the central bar and spiral arms.

We note here a few of the many important and challenging science cases,
all of which require GAIA's faint limiting magnitude, and which
illustrate GAIA's study of the Galactic Bulge, Disk, and Halo.

\subsection{The Galactic Bulge}

Bulge stars are predominantly moderately old, unlike the present-day
disk; they encompass a wide abundance range, peaking near the Solar
value, as does the disk; and they have very low specific angular
momentum, similar to stars in the halo. Thus the bulge is, in some
fundamental parameters, unlike both disk and halo. What is its
history? Is it a remnant of a disk instability? Is it a successor or a
precursor to the stellar halo? Is it a merger remnant?  It is not
clear whether the formation of the bulge preceded that of the disk, as
predicted by `inside-out' scenarios; or whether it happened
simultaneously with the formation of the disk, by accretion of dwarf
galaxies; or whether it followed the formation of the disk, as a
result of the dynamical evolution of a bar.  Large-scale surveys of
proper motions and photometric data inside the bulge can cast light on
the orbital distribution function. Knowing the distance, the true
space velocities and orbits can be derived, thus providing constraints
on current dynamical theories of formation. GAIA data for bulge stars,
providing intrinsic luminosities, metallicity, and numbers, can be
inverted to deduce star formation histories. 

The highly accurate parallaxes, proper motions and magnitudes acquired
by GAIA for more than $10^6$ stars per square degree, will allow the
vast majority of red and asymptotic giant branch stars, and a
significant fraction of the clump stars in Baade's Window to be
measured with a precision higher than 10--15~per cent. With $V=20$ as
the limiting magnitude, red and asymptotic giant branch stars can be
detected over a range of 5~mag.

There is substantial evidence that the bulge is not axisymmetric, but
instead has a triaxial shape seen nearly end-on.  Indications for this
come from the asymmetric near-infrared light distribution, star
counts, the atomic and molecular gas morphology and kinematics, and
the large optical depth to micro-lensing. The actual shape, orientation, and
scale-length of the bulge, and the possible presence of an additional
bar-like structure in the disk plane, however remain a matter of debate. 
The reason why it is so difficult to derive the shape
of the Galactic bar is that three-dimensional distributions cannot be
uniquely recovered from projected surface brightness distributions
such as the COBE/DIRBE maps. In addition, bars with the same density
distribution could have different pattern speeds.  No unique solution
can be found using only one-velocity component diagrams, unless the
gravitational potential is known, since the velocity dispersion in the
star motions smears out the effects of the bar on the distribution
function.

GAIA proper motions to faint magnitudes, in particular in a number of
low-extinction windows, will allow unambiguous determination of the
shape, orientation, tumbling rate mass profile and star formation
history of the bulge. The large-scale kinematics of the Galaxy also
contains an imprint of the non-axisymmetric central potential.

\subsection{The Galactic Halo}

The stellar halo of the Galaxy contains only a small fraction of its
total luminous mass, but the kinematics and abundances of halo stars,
globular clusters, and the dwarf satellites contain imprints of the
formation of the entire Milky Way.  The most metal-deficient stars,
with [Fe/H]~$< -3.5$, represent a powerful tool to understand
primordial abundances and the nature of the objects which produced the
first heavy elements.

\paragraph{Halo Streams} The halo of the Milky Way is likely to be the
most important component that may be used to distinguish among
competing scenarios for the formation of our Galaxy. The classical
picture of inner monolithic collapse with later accretion in the outer
Galaxy, predicts a smooth distribution both in configuration and
velocity space for our Solar neighbourhood, which is consistent with
the available observational data. The currently popular hierarchical
cosmologies propose that big galaxies are formed by mergers and
accretion of smaller building blocks, and many of its predictions seem
to be confirmed in high-redshift studies.

Those merging and accretion events leave signatures in the phase-space
distribution of the stars that once formed those systems.
Helmi etal have shown that, after
10~billion years, the spatial distribution of stars in the inner halo
should be fairly uniform, whereas strong clumping is expected in
velocity space. This clumping appears in the form of a very large
number of moving groups (several hundred in a 1~kpc$^3$ volume
centered on the Sun, if the whole stellar halo were built
in this way) each having very small velocity dispersions and
containing several hundred stars.  The required velocity accuracies to
detect individual streams are less than a few~km~s$^{-1}$, requiring
measurement precision of order $\mu as$.

\paragraph{The Outer Halo} GAIA will find several million individual 
stars in the outer halo (defined here as galactocentric distance
$R>20$~kpc). These will mostly be G and K giants and red and blue
horizontal branch stars. G and K giants are intrinsically bright, they
form in all known old stellar population types, they have easily
measurable radial velocities, and they are historically well studied
because they are the most easily accessible stars in the globular
clusters. Horizontal branch stars have been the preferred tracer
stellar type for the outer halo to date, because they can be much more
easily identified amongst field stars than G and K giants. In
particular, blue horizontal branch stars have been very easy to
locate, since almost all faint ($14<V<19$~mag), blue $(0.0<B-V<0.2$)
stars are halo blue horizontal branch stars.  However
these stars are a biased tracer of the halo population in the sense
that they do not always form in old metal weak populations (viz.\ the
second parameter problem in globular clusters). Redder horizontal
branch stars and G and K halo giants are drowned out by the huge
numbers of foreground turnoff and dwarf stars in the Galactic disk.

GAIA will circumvent all these difficulties. The late-type foreground
dwarfs are much closer than the background late-type giants, so that
at faint magnitudes ($V<19$~mag) the dwarfs have a measurable parallax
while the background giants do not. It will be possible to lift the
veil of foreground stars and reveal of order millions of background
halo stars, on the giant branch, and the red and blue horizontal
branch.

\subsection{Large Scale Structure of the Galactic Disk}

We note here just two of the many aspects of this science GAIA will
 explore. 

\paragraph{Galactic Disk Warps}
Galactic disks are thin, but they are not flat.  Approximately
one-half of all spiral galaxies have disks which warp significantly
out of the plane defined by the inner galaxy.  Remarkably, there is no
realistic explanation of this common phenomenon, though the
large-scale structure of the dark matter, and tidal interactions, must
be important, as the local potential at the warp must be implicated.
Neither the origin nor the persistence of galaxy warps is understood,
and insufficient information exists to define empirically the relative
spatial and kinematic distributions of the young (OB) stars which
should trace the gas distribution, and the older (gKM) stars which
define a more time-averaged gravitational field.

The expected kinematic pattern (at least, in existing
plausible models) is most strongly constrained by the straightness of
the line of nodes: these should wind up in at most a few rotation
times, typically less than 2~Gyr. A relevant shear pattern corresponds
to systematic motions dependant on warp phase and galactocentric
distance superimposed on Galactic rotation.
A plausible velocity amplitude associated with the warp at the optical
disk edge is significantly less than $0.1\Omega$, with $\Omega$ the
disk rotation angular velocity. This will be distributed between
latitude and longitude contributions depending on the local geometric
projection.

At $R=15$~kpc, for a flat rotation curve, the systematic disk rotation
corresponds to 6~mas~yr$^{-1}$. The kinematic signature from a
1~kpc-high warp corresponds to a systematic effect of
$\sim90$~$\mu$as~yr$^{-1}$ in latitude and $\sim600$~$\mu$as~yr$^{-1}$
in longitude. For such a signal to be detected the reference frame
must be rigid to better than a few microarcsec on scales of
$\sim10^\circ$ (i.e.\ matching the high-frequency warp structure) and
on scales of $2\pi$~radians, requirements well within the GAIA
capabilities. The corresponding distance requirements are more
demanding: at the warp a mean parallax is less than 100~$\mu$as, so
that resolution of the warp within 10~per cent implies distance
accuracies of 10~$\mu$as at $I\sim15$~mag. Along lines of sight with
typical reddening, the study of the Galactic warp will be within the
limits of GAIA's performance. 

\paragraph{Dark Matter in the Disk} The distribution of mass in the 
Galactic disk is characterized by two numbers, its local volume
density $\rho_o$ and its total surface density $\Sigma(\infty)$.
They are fundamental parameters for many aspects of Galactic
structure, such as chemical evolution (is there a significant
population of white dwarf remnants from early episodes of massive star
formation?), the physics of star formation (how many brown dwarfs are
there?), disk galaxy stability (how important dynamically is the
self-gravity of the disk?), the properties of dark matter (does the
Galaxy contain dissipational dark matter, which may be
fundamentally different in nature from the dark matter assumed to
provide flat rotation curves, and what is the local dark matter
density and velocity distribution expected in astroparticle physics
experiments?), and non-Newtonian gravity theories (where does a
description of galaxies with non-Newtonian gravity and no dark matter
fail?).
 
The most widely referenced and commonly determined measure of the
distribution of mass in the Galactic disk near the Sun is the local
volume mass density $\rho_o$, i.e.\ the amount of mass per unit
volume near the Sun, which for practical purposes is the same as the
volume mass density at the Galactic plane.  This quantity has units of
$M_\odot$~pc$^{-3}$, and its local value is often called the `Oort
limit'. The contribution of identified material to the Oort
limit may be determined by summing all local observed matter -- an
observationally difficult task.  The uncertainties arise in part due
to difficulties in detecting very low luminosity stars, even very near
the Sun, in part from uncertainties in the binary fraction among low
mass stars, and in part from uncertainties in the stellar
mass--luminosity relation. All these quantities will be determined
directly, to extremely high precision, by GAIA.

The second measure of the distribution of mass in the Solar vicinity
is the integral surface mass density. This quantity has units of
$M_\odot$~pc$^{-2}$, and is the total amount of disk mass
in a column perpendicular to the Galactic plane. It is this quantity
which is required for the deconvolution of rotation curves into `disk'
and `halo' contributions to the large-scale distribution of mass in
galaxies. If one knew both the local $\rho_o$ and $\Sigma(\infty)$,
one could immediately constrain the scale height of any contribution
to the local volume mass density which was not identified.  That is,
one could measure directly the velocity dispersion, i.e., the
temperature, of the `cold' dark matter.

\subsection{Stellar Astrophysics}

GAIA will provide distances of delightful accuracy for all types of
stars of all stellar populations, even the brightest, or those in the
most rapid evolutionary phases which are very sparsely represented in
the Solar neighbourhood.  With the parallel determination of
extinction/reddening and metallicities by the use of multi-band
photometry and spectroscopy, this huge amount of basic data will
provide an extended basis for reading {\it in situ\/} stellar and galactic
evolution.  All parts of the Hertzsprung--Russell diagram will be
comprehensively calibrated, including all phases of stellar evolution,
from pre-main sequence stars to white dwarfs and all existing
transient phases; all possible masses, from brown dwarfs to the most
massive O~stars; all types of variable stars; all possible types of
binary systems down to brown dwarf and planetary systems; all standard
distance indicators (pulsating stars, cluster sequences, supergiants,
central stars of planetary nebulae, etc.). This extensive amount of
data of extreme accuracy will stimulate a revolution in the
exploration of stellar and Galactic formation and evolution, and the
determination of the cosmic distance scale.

The agreement between predicted and observed properties of stars has
remained qualitative due to the modest accuracy and relative scarcity
of the relevant observed quantities.  Luminosity measurements are
based exclusively on determinations of stellar distances and
interstellar absorption. Absorption can be deduced from multi-colour
photometry, obtainable with GAIA. The distances can be determined
directly only by measurement of the trigonometric parallax. GAIA will
provide distances to an unprecedented 0.1~per~cent for $7\times 10^5$
stars out to a few hundred pc, and to 1~per~cent accuracy for a
staggering $2.1\times 10^7$ stars up to a few kpc. Distances to
10~per~cent will reach beyond 10~kpc, and will cover a significant
fraction of our Galaxy, including the Galactic centre, spiral arms,
the halo, and the bulge, and---for the brightest stars---to the
nearest satellites.  The faint limiting magnitude allows investigation
of white dwarfs as well as the bottom of the main sequence down to
brown dwarfs.  For the first time, this will provide an extensive
network of accurate distance measurements for all stellar types. 

The ability to determine simultaneously and systematically the
planetary frequency and distribution of orbital parameters for the
stellar mix in the Solar neighbourhood is a fundamental contribution
that GAIA will uniquely provide. The only limitations are those
intrinsic to the mission, i.e., the actual sensitivity of the GAIA
measurements to planetary perturbations.  GAIA's strength will be its
discovery potential, following from the combined photometric and
astrometric monitoring of all of the several hundred thousand bright
stars out to distances of $\sim200$~pc.

Essentially all Jupiter-mass planets within 50~pc and with periods between
1.5--9~years will be discovered by GAIA. 

\subsection{Solar System}

Solar system objects present a challenge to GAIA because of their
significant proper motions, but they promise a rich scientific
reward. The minor bodies provide a record of the conditions in the
proto-Solar nebula, and their properties therefore shed light on the
formation of planetary systems. Discovery and orbital determination of
near-Earth objects is a subject of high public interest.

In addition to known asteroids, GAIA will discover a very large
number, of the order of $10^5$ or $10^6$ (depending on the
uncertainties on the extrapolations of the known population) new
objects.  It should be possible to derive precise orbits for all the
newly discovered objects, since each of them will be necessarily
observed many times during the mission lifetime.  These will include a
large number of near-Earth objects.  GAIA is ideal to look for these
objects because of the enormous area of sky that must be searched.

GAIA will detect a significant number of Kuiper Belt objects during
its 5-year mission. The angular motion of a typical object at
$\sim90^\circ$ elongation (where GAIA will be looking) is small: the
known KBOs have $d\alpha/dt=0.02-1.0$~arcsec~hr$^{-1}$ and
$d\delta/dt=0.002-1.2$~arcsec~hr$^{-1}$. The surface density of the
Kuiper Belt at $V=20$~mag is $8 \times 10^{-3}$ objects per square
degree ($2\times10^{-2}$ at $V=21$~mag, implying that GAIA should
discover some number up to $\sim300$~KBOs with $V\le20$ ($\sim800$
KBOs with $V\le21$).

Scientific objectives regarding the Kuiper Belt that can be answered
only with GAIA include binarism, new Plutinos, and the good orbits
essential to understand the system dynamics.

\subsection{The Local Group, distant Galaxies, Quasars, and the Reference Frame}

GAIA will not only provide a representative census of the stars
throughout the Milky Way, but it will also make unique contributions
to extragalactic astronomy. These include the structure, dynamics and
stellar populations in the Magellanic Clouds and other Galactic
satellites, and in M31 and M33, with scientific consequences
comparable to those noted above for the Milky Way.  In addition, the
faint magnitude limit and all-sky survey of GAIA allows unique
cosmological studies, from the space motions of Local Group galaxies,
and studies of huge numbers of supernovae, galactic nuclei, and
quasars.

\paragraph{Orbits in the Local Group: Gravitational Instability 
                in the Early Universe}
The orbits of galaxies are a result of mildly non-linear gravitational
interactions, which link the present positions and velocities to the
cosmological initial conditions. Non-gravitational (hydrodynamic) or
strongly non-linear gravitational interactions (collisions, mergers)
are sometimes significant. It is uniquely possible in the Local Group
to determine reliable three-dimensional orbits for a significant
sample of galaxies, in a region large and massive enough to provide a
fair probe of the mass density in the Universe.  Such orbital
information provides direct constraints on the initial spectrum of
perturbations in the early Universe, on the global cosmological
density parameter $\Omega$, and on the relative distributions of mass
and light on length scales up to 1~Mpc.

Radial velocities are known.  The required measurements are distances
and transverse velocities for the relatively isolated members of the
Local Group, those more distant than $\sim$100~kpc from another large
galaxy.  Improved distances will be derived from the GAIA-calibrated
standard distance indicators, such as Cepheids and RR~Lyraes.  The
transverse motions will be derivable uniquely from the GAIA proper
motion.

\paragraph{Galaxies, Quasars and Supernovae:}

Growth of structure in the Universe is believed to proceed from small
amplitude perturbations at very early times. Growth from the
radiation-dominated era to the present has been extensively studied,
particularly in the context of the popular hierarchical clustering
scenario. Many aspects of this picture are well-established. Others
are the subject of active definition through redshift and imaging
surveys of galaxies, and the microwave background experiments. There
are several aspects of this research which require very wide area
imaging surveys with high spatial resolution, to provide
high-reliability catalogues of galaxies and quasars extending to low
Galactic latitudes.  Here GAIA will contribute uniquely, by detecting
and providing multi-colour photometry with $\sim$0.3~arcsec spatial
resolution for all sufficiently high-surface brightness galaxies.
This provides a valuable and unique data set at two levels: for
statistical analysis of the photometric structure of the central
regions of many tens of thousands of galaxies; and for study of the
large-scale structure of the local Universe. The scientific value of
this huge and homogeneous database will impact all fields of galaxy
research, naturally complementing the several redshift surveys, and
the deeper pencil-beam studies with very large telescopes. Among the
most important unique GAIA science products will be determination of
the colour and photometric structure in the central regions of a
complete, magnitude-limited sample of relatively bright galaxies.

{ Supernovae:}
GAIA will detect all compact objects brighter than V$=20$~mag, so that in
principle supernovae can be detected to a modulus of $m-M\sim39$~mag,
i.e., to a distance of 500~Mpc or $z\sim0.10$. Simulations show that
in 4~years, GAIA will detect about $100\,000$ supernovae of all types.
Of these, the most useful as cosmological-scale distance indicators
are the Type~Ia supernovae, whose light curves are very accurate
distance indicators, $\pm 5$ per cent. Rapid detection of such
transient sources will allow detailed ground-based determination of
lightcurves and redshifts.

{ Quasars:} The astrometric programme to $V=20$~mag will provide a
census of $\sim500\,000$ quasars.  The mean surface density of $\sim
25\,{\rm deg}^{-2}$ at intermediate to high Galactic latitudes will
provide the direct link between the GAIA astrometric reference system
and an inertial frame. They are also of direct astrophysical interest.

Existing ground-based studies of gravitational (macro) lensing among
the quasar population are restricted to resolutions of $\sim1$~arcsec.
 GAIA will provide sensitivity to multiply-imaged systems with
separations as small as $\sim0.2$~arcsec.  For the brighter quasars,
$V<18$~mag, with a surface density of $\sim1~{\rm deg}^{-2}$, where
examples of lensing are most common, GAIA's sample of $\sim 50\,000$
quasars represents an increase of two orders of magnitude over
existing surveys. Pushing the sensitivity to image separations of a
few tenths of an arcsec will access systems where most of the lensing
due to individual galaxies is expected. In particular, the GAIA survey
will provide new constraints on lensing by the bulk of the galaxy
population, including spiral galaxies, rather than the high-mass tail
of ellipticals to which existing surveys are predominantly sensitive.
This homogeneous sample would provide decisive astrophysical
information, including constraints on the cosmological parameters 
$\Omega$ and $\lambda_0$. Photometric variability of  multiply
lensed quasars is of course also a proven method to determine ${\rm H_o}$.

\paragraph{Reference Frames} 
At present, the International Celestial Reference System (ICRS) is
primarily realized by the International Celestial Reference Frame
(ICRF) consisting of positions of 212~extragalactic radio-sources
with an rms uncertainty in position between 100 and
500~$\mu$as. The extension of the ICRF to visible light is the
Hipparcos Catalogue with rms uncertainties estimated to be
0.25~mas~yr$^{-1}$ in each component of the spin vector of the frame
($\omega$) and 0.6~mas in the components of the orientation vector
($\varepsilon$) at the catalogue epoch, J1991.25.
The GAIA catalogue will permit a definition of the ICRS more
accurate by three orders of magnitude than the present
realizations. GAIA will define the ICRS to better than
60$\mu$as in the orientation of the frame.

\subsection{Fundamental Physics: The Space-Time metric}

The dominating relativistic effect in the GAIA measurements is
gravitational light bending. Accurate measurement of the parameter
$\gamma$ of the Parametrized Post-Newtonian (PPN) formulation of
gravitational theories is of key importance in fundamental physics.
The Pound-Rebka experiment verified the relativistic prediction of a
gravitational redshift for photons, an effect probing the time-time
component of the metric tensor. Light deflection depends on both the
time-space and space-space components. It has been observed, with
various degrees of precision, on distance scales of $10^9-10^{21}$~m,
and on mass scales from $1-10^{13} M_\odot$, the upper ranges
determined from the gravitational lensing of quasars.  GAIA will
extend the domain of observations by two orders of magnitude in
length, and six orders of magnitude in mass.

Detailed analyses indicate that the GAIA measurements will provide a
precision of about $5\times10^{-7}$ for $\gamma$. This accuracy is
close to the values predicted by theories which predict that the
Universe started with a strong scalar component, which relaxes to the
general relativistic value with time.

\section{GAIA: the mission}

GAIA will be a continuously scanning spacecraft, accurately
measuring one-dimensional coordinates along great circles, and in two
simultaneous fields of view, separated by a well-defined and
well-known angle. These one-dimensional coordinates are then converted
into the astrometric parameters in a global data analysis, in which 
distances and proper motions `fall out' of the processing, as does
information on double and multiple systems, photometry, variability,
metric, planetary systems, etc. The payload is based on a large but 
feasible CCD focal plane assembly, with passive thermal control, and a
natural short-term (3~hour) instrument stability due to the sunshield,
the selected orbit, and a robust payload design.

The telescopes are of moderate size, with no specific design or
manufacturing complexity. The system fits within a dual-launch Ariane
5 configuration, without deployment of any payload elements. A
`Lissajous' orbit at the outer Lagrange point L2 has been identified
as the preferred operational orbit, from where an average of 1~Mbit of
data per second is returned to the single ground station throughout
the 5-year mission. The 10~microarcsec accuracy target has been shown
to be realistic through a comprehensive accuracy assessment programme;
this remarkable accuracy is possible partly by virtue of the (unusual)
instrumental self-calibration achieved through the data analysis
on-ground. This ensures that final accuracies essentially reflect the
photon noise limit for localisation accuracy: this challenge, while
demanding, has been proven deliverable by the Hipparcos experience.

\subsection{GAIA: the observatory}

GAIA will record more than just huge volumes of positional data on a
vast number of astrophysical targets. GAIA will also provide a
complementary range of data, with a diversity of applications. Every
one of the 10$^9$ GAIA targets will be observed typically 100 times,
each time in a complementary set of photometric filters, and a large
fraction also with a radial velocity spectrograph. The available
spatial resolution exceeds that available in ground-based
surveys. Source detection happens on-board at each focal-plane
transit, so that variable and transient sources are detected. All
these complementary datasets, in addition to the superb positional and
kinematic accuracy which is derivable from their sum, make GAIA an
optimal observatory mission: every observable source will be observed
every time it crosses the focal plane.

These data allow studies from asteroids to distant supernovae, from
planets to galaxies, and naturally interest almost the entire
astronomical community. Because of this enormous interest, GAIA will
be an open observatory mission, directly making available its rich
scientific resource to the sponsoring communities.  The scale of the
GAIA data is such that many analyses can be undertaken during
operations, some will require the whole mission calibration
information, while others again will await final data reduction.  The
GAIA observatory will provide exciting scientific data to a very wide
community, beginning with the first photometric observations, and
rapidly increasing until the fully reduced GAIA data become
available. The resulting analyses will provide a vast scientific
legacy, providing a wealth of quantitative data on which all of
astrophysics will build.

\subsection{GAIA Mission Organisational Structure}

GAIA is unusual for a space project, in that ESA will provide the
payload inside the mission budget. This results in vastly less
pressure on national science budgets, but presumes a slightly unusual
project management structure.

Following selection of GAIA as Cornerstone~6, the work of the ESA
Science Advisory Group was complete. A new structure is currently
being implemented, to guide the project over the next few years
towards final detailed definition, and into construction. The ESA
effort is supervised by a project scientist (Michael Perryman,
formerly HIPPARCOS project scientist) and a project manager
(Oscar Pace, the GAIA study manager). Two Industrial contracts are
soon to be placed to oversee system development, and the many
technical studies underway in European industry. These studies will be
overseen by a GAIA Science and Technical Advisory Group, chaired by
Michael Perryman. The members of this act as coordinators for the
(currently 20) groups of community scientists who are active in
defining GAIA. A Space Astrometry Forum (Chair: Francois Mignard,
Nice) brings together the various space astrometry missions for issues
such as reference frames. General oversight of the GAIA studies is
provided by the GAIA Consultative Committee (Chair: Gerry Gilmore).

\subsection{ GAIA: How can I be involved?}

There are many scientific tasks to optimise GAIA which demand
immediate effort. GAIA will be defined and designed beyond major
change in 3 years time. We have that much time to get it right.  This
challenge provides an ideal opportunity for interested scientists, and
for PhD projects, to play a major role in the project, and in the
future of astronomy.  Among these tasks are definition of the GAIA
photometric system, development of data reduction algorithms and
methods, modelling the sky as seen by GAIA, developing systems for
real-time access to GAIA photometry, and so on. The GAIA data will
underpin astronomy for decades to come. The techniques needed to
analyse 100Tb of multi-dimensional data are those of relevance for all
future research. The science content of 100Tb of GAIA data is vast.
Work has started. Much needs to be done.

Scientists interested in being part of GAIA activities, or wanting to
know more, should contact Michael Perryman 
(mperryma@astro.estec.esa.nl) or Gerry Gilmore (gil@ast.cam.ac.uk).

\section{Conclusion}

GAIA addresses science of vast general appeal, and will deliver huge
scientific impact across the whole of astrophysics from studies of the
Solar System, and other planetary systems, through stellar
astrophysics, to its primary goal, the origin and evolution of
galaxies, out to the large scale structure of the Universe, and
fundamental physics.

GAIA is timely as it builds on recent intellectual and technological
breakthroughs. Current understanding and exploration of the early
Universe, through microwave background studies (e.g., Planck) and
direct observations of high-redshift galaxies (HST, NGST, VLT) have
been complemented by theoretical advances in understanding the growth
of structure from the early Universe up to galaxy formation. Serious
further advances require a detailed understanding of a `typical'
galaxy, to test the physics and assumptions in the models. The Milky
Way and the nearest Local Group galaxies uniquely provide such a
template.

While challenging, the entire GAIA design is within the projected
state-of-the-art, and the satellite can be developed in time for
launch in 2010. With such a schedule, a complete stereoscopic map of
our Galaxy will be available within 15~years.  GAIA will provide a
quantitative, stereoscopic movie of the Milky Way, and so unlock its
origins.

\end{document}